\documentclass[a4paper,fleqn]{cas-sc}
\usepackage{amsthm,amsmath}
\usepackage[linesnumbered,ruled]{algorithm2e}

\usepackage[numbers]{natbib}

\def\tsc#1{\csdef{#1}{\textsc{\lowercase{#1}}\xspace}}
\tsc{WGM}
\tsc{QE}

\begin{document}
\let\WriteBookmarks\relax
\def\floatpagepagefraction{1}
\def\textpagefraction{.001}

\shorttitle{A Microgrid Trading Framework Based on PoC Consensus}    

\shortauthors{Lianghaojie Zhou et al.}  

\title [mode = title]{A Microgrid Trading Framework Based on PoC Consensus}  



%

\author{Lianghaojie Zhou}[
type=editor,
orcid=0009-0008-8529-930X
]


\ead{zhoulianghaojie@stu.gxnu.cn}



\affiliation{organization={Guangxi Normal University},
                city={GuiLin}, 
                country={China}}

\author{Youquan Xian}


\ead{xianyouquan@stu.gxnu.edu.cn}

\author{Yipeng Yang}
\ead{yyp@stu.gxnu.edu.cn}

\author{Jianyong Jang}
\ead{jiangyong@stu.gxnu.edu.cn}

\author{Peng Liu}
\ead{liupeng@gxnu.edu.cn}

\author{Xianxian Li}
\ead{lixx@gxnu.edu.cn}
\cormark[1]



\cortext[1]{Corresponding author. College of Computer Science and Engineering,Guangxi Normal University,GuiLin,China.}


\begin{abstract}
 In the field of energy Internet, blockchain-based distributed energy trading mode is a promising way to replace the traditional centralized trading mode. However, the current power blockchain platform based on public chain has problems such as low consensus efficiency and waste of computing resources. The energy trading platform based on the consortium chain has problems such as inability to attract users to join and collusion between alliances. We propose a microgrid trading framework based on proof of contribution (PoC). According to the contribution value, we randomly select nodes based on weights through verifiable random functions (VRF) in each round of consensus to form the next round of consensus committee. The proposed electricity trading framework not only  improves the consensus efficiency of the blockchain, but also is suitable as an incentive mechanism to attract users to participate in the power blockchain. Experiments show that our framework is effective and the contribution value we designed is reasonable. Through our framework, we can motivate users to participate and make energy transactions more fair.
\end{abstract}




\maketitle

\section{Introduction}\label{}
With the increasing maturity of distributed power generation equipment and renewable energy technology, the concept of energy internet has become the mainstream of future development. At the same time, the traditional power grid production and marketing model has problems such as low efficiency, opaque transactions, and vulnerability to single-node attacks. The emergence of Peer-to-Peer (P2P) trading model allows consumers and producers to achieve 'face-to-face' transactions. Because there is no central server, P2P transactions not only improve the efficiency of the transaction, but also increase the income of both parties. The combination of de-neutralized blockchain technology and P2P transactions have brought new opportunities and challenges to energy transactions\cite{esmat2021novel}. Blockchain\cite{2008Bitcoin} technology is a distributed ledger proposed in 2008, which guarantees the non-tamperability of data and prevents data fraud through cryptography. Blockchain is mainly used to solve the trust and security problems of transactions. The emergence of smart contracts\cite{ethereum2014ethereum} allows blockchain to be applied in more scenarios. Smart contracts are programmable codes that run on the blockchain. Once the predefined conditions in the contract are met, the program will execute automatically. We take the energy trading of microgrid as an example. Electricity transaction information and electricity data are stored on the blockchain, and the settlement of energy transactions can be automatically completed  through smart contracts\cite{kirli2022smart}.

However, in the scheme of energy trading based on blockchain, the traditional consensus methods are not suitable. Since the public chain consensus algorithm relies on cryptocurrency, the consensus efficiency is low, and the consortium chain lacks corresponding incentive mechanism. In this paper, We propose a micro-grid trading framework based on contribution consensus. We quantify the contribution value of nodes based on their on-chain and off-chain behavior, perform energy trading and blockchain consensus based on the contribution value. We use the  VRF algorithm based on contribution value to randomly select nodes to form a formula committee to improve the efficiency and fairness of the consensus. The main contributions of this paper are summarized below:

\begin{itemize}
  \item [1)] 
We propose a blockchain microgrid electricity trading framework based on PoC consensus. The framework does not rely on cryptocurrency and certificate authorities, but use contribution value as an incentive to attract users to actively participate in energy trading.
  \item [2)]
We propose a new PoC consensus process that uses VRF to randomly select consensus nodes based on contribution value weights. We divide the contribution value into transaction contribution value and consensus contribution value, and only selected nodes can perform contribution value calculation.
  \item [3)]Finally, we evaluate the security and performance of PoC consensus and compare it with the traditional PoW consensus. The results show that our proposed system is indeed efficient
\end{itemize}

 The rest of this paper is organized as follow: Section 2 introduces some related work on consensus and energy trading framework. Section 3 introduces the energy trading system model, and the P2P electricity trading process. Section 4 introduces the PoC consensus process. Section 5 introduces the algorithm for calculating contribution values. Section 6 conducts a security analysis of our framework. Section 7 evaluates our consensus algorithm and compares it with PoW. Finally, Section 8 summarizes the paper and discusses some future work.

\section{Related work}\label{}
    In the current consensus algorithms used by public blockchains, Proof of Work (PoW)\cite{dwork1992pricing} uses a puzzle-solving mechanism to ensure the credibility of blocks. This puzzle is typically a computationally difficult but easy-to-verify problem. But PoW needs to spend a lot of computing resources to calculate these meaningless puzzles, which is not economical and environmentally friendly\cite{zheng2018blockchain}. In comparison with PoW, proof-of stake (PoS)\cite{king2012ppcoin} can be an energy efficient alternativehas been proposed. PoS is an efficient cryptocurrency, which is based on coin age. Coin age is simply defined as the number of currencies multiplied by the holding period. The more coin age you haved, the easier findig the hash target. But the PoS is prone to Matthew effect and centralization. The consortium chain solves the problem of resource waste in the public chain by raising the access threshold, but at the same time, there are problems such as node collusion and lack of corresponding incentive mechanism. 

    In order to solve the above problems, consensus algorithms based on contribution value were proposed. Hongyu Song et al.\cite{song2021proof} proposed a consensus algorithm based on contribution value and applied it in the field of intellectual property (IP) rights. The node with the highest contribution value is selected as the accounting node. Jingyu Feng et al.\cite{feng2020towards} proposed a method of randomly selecting honest miners. They set up a trusted third-party reputation management agency to manage the reputation value of nodes in the network, and let the reputation management agency perform random selection algorithm to select qualified miners as proposal nodes and verification nodes. Wenjun Cai et al.\cite{cai2020dynamic} proposed a dynamic reputation PBFT consensus algorithm for energy blockchain. They divide nodes into different roles based on reputation values, allowing them to participate in different stages of consensus. Adamu Sani Yahaya et al.\cite{9276408} proposed a reputation-based workload proof consensus (PoWR), which can effectively reduce the time required for transaction confirmation and block creation. Li-e Wang et al.\cite{wang2020beh} proposed a blockchain protocol by combining Proof-of-Behavior (PoB) with Raft, which reduces the time complexity of consensus and ensures the security of consensus process. Through the above research, it can be found that the consensus algorithm based on contribution value can improve the efficiency of blockchain consensus. But the management of reputation value becomes a problem. The third party management reputation value will be too centralized. Let node management will allow nodes with high reputation values to be maliciously attacked.

    Meantime, the combination of blockchain technology and energy trading brings a new trading paradigm. \cite{luo2018distributed}, \cite{huang2022two},\cite{esmat2021novel} ,\cite{zhang2021peer} proposed blockchain trading platforms to maximize the interests of users and improves the efficiency of transactions through layered design, game theory and double auction. YuTian Lei et al.\cite{lei2022renewable} proposed a homomorphic encryption scheme to provide privacy protection for transaction data. Keke Gai et al.\cite{gai2019privacy} propose to use multiple anonymous accounts in the transaction process to protect the user's real information. However, due to the random fluctuations in energy prices, some malicious users will deliberately default to seek more benefits. The above research do not consider the credibility of energy trading users.
    
    Kaile Zhou et al.\cite{9531949} proposed a P2P electricity trading model considering user reputation. By setting different waiting times for users with different reputation levels, users with high reputation values are preferentially involved in transactions. Tonghe Wang et al.\cite{wang2021rbt} automatically manage reputation values through smart contracts. They designed a reputation-based PBFT consensus. Only nodes with high reputation values participate in the consensus, and other participants only passively receive messages. Zhitao Guan et al.\cite{guan2021achieving}  proposed a reputation-based PoS. Each round of consensus selects the node with the highest reputation value as the consensus node, and the remaining nodes participate in the verification of the packaged block. Longze Wang et al.\cite{wang2021credible} would reward or punish users according to their behavior in the energy trading phase. Users with higher credit can act as notaries and participate in the verification of system information. Bilal Ahmad Bhatti\cite{bhatti2019energy} quantified users' behavior as an market reputation index (MRI). The higher the MRI, the higher the transaction price. Zhiyi Li et al.\cite{li2019blockchain} believed that users with low reputation values are selfish nodes and will be restricted from participating in energy transactions. The above studies are not friendly to the recently added users. If there is no measure, some users who joined the trading platform in the early days, or large enterprises will cause a monopoly of reputation, so they will always enjoy the benefits.
    
    We believe that the reputation/contribution value should not be regarded as a measure of the price or order of the transaction. Or directly selects the node with the highest contribution value as the consensus node of this round. Because of the transparent nature of the blockchain, malicious nodes can guess which node has the highest contribution value, which is unsafe for the entire blockchain. If privacy protection schemes such as homomorphic encryption and attribute encryption are adopted, the efficiency of blockchain consensus will be further reduced. To solve the above problems, We proposed a microgrid electricity trading framework based on improved PoC consensus. We classify nodes and contribution values. Only selected nodes can calculate the contribution value. Based on the weight of the credibility value, the members of the consensus committee for the next round are randomly selected. In the electricity trading stage, users with high contribution values have priority trading rights.

    \section{Energy trading framework based on the PoC}\label{}
      \begin{figure}[h!]
      \centerline{   \includegraphics[width=8cm,height=6cm]{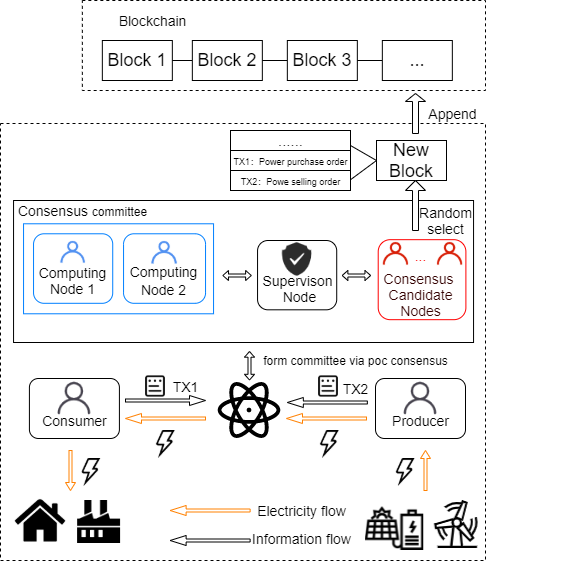}}
  \caption{The proposed energy trading framework.}
      \label{fig:my_label}
\end{figure}

In this section, we will introduce the energy trading framework based on the PoC consensus. The energy trading system model is shown in Figure 1.
We divide nodes in the blockchain network into four roles : computing node ($N_{cp}$), consensus candidate node ($N_{cs}$), ordinary node ($N_{o}$) and supervision node($N_{s}$).
\begin{itemize}
  \item [1)]Computing node: It is responsible for calculating the contribution of nodes in the blockchain network during the previous round of consensus. It executes a random selection algorithm based on contribution value weights to select consensus candidate and computing node for the next round.
  \item [2)]Consensus candidate node: Responsible for packaging and chunking the received transactions during the round of consensus. Consensus node is randomly selected from candidates through VRF. In a round of consensus, the computing node and the consensus candidates work simultaneously. 
  \item [3)]Ordinary node: Responsible for forwarding transactions and verifying this round of block. All nodes are ordinary nodes at the beginning.
  \item [4)]Supervision node: As a special node in the blockchain network, $N_{s}$ is responsible for daily maintenance by the power grid company and can be considered as a trusted node. $N_{s}$ is responsible for generating random number seeds and forwarding messages honestly.
  \end{itemize}

   As shown in the system model, users with green distributed power generation equipment participate in the blockchain network as nodes of the blockchain. A simple summary is user A and B submit orders (purchase order or sale order) to the blockchain platform. Power transactions are automatically settled through smart contracts. The specific electricity trading process is described later. When $N_{cs}$s package the current round of transactions, the $N_{cp}$s are calculating the contribution value of the nodes in the previous round of consensus rounds, and randomly selects the next round of candidate nodes. $N_{s}$ verifies the behavior of $N_{cp}$s and the feasibility of the candidates. Finally, when the next round of committees participating in the consensus are all elected, the out-block nodes of this round will be randomly selected by the VRF method.
  
Our blockchain energy trading processes involves four steps: order generation, order matching, order execution, and order settlement.The specific energy trading process is shown in Figure 2. The trading platform is responsible for contribution value calculation, order matching, transaction records and order settlement. Users are responsible for submitting orders, selecting orders, and other operations.

In the order generation phase, there are already mature machine learning methods\cite{9144528} for accurate power load forecasting of distributed generation energy sources. Users submit their orders to the blockchain trading platform based on the forecast results and must ensure that there are sufficient funds in their account balance.

In the order matching phase, the trading platform matches transactions based on the user's energy trading contribution value. When supply exceeds demand, seller orders are ranked in descending order by contribution value, and seller orders with high contribution values are matched first. The buyer and seller confirm the information of the order, which includes the quantity of electricity traded, the unit price of the transaction, and the latest delivery time.

In the order execution phase, the generating party shall issue sufficient power by the specified latest time as specified in the order, and then dispatch the power.

In the order settlement stage, the trading platform settles electricity bills through smart contracts according to the actual electricity delivered.

\begin{figure}[h]
    \centering
    \includegraphics[width=8.5cm,height=6cm]{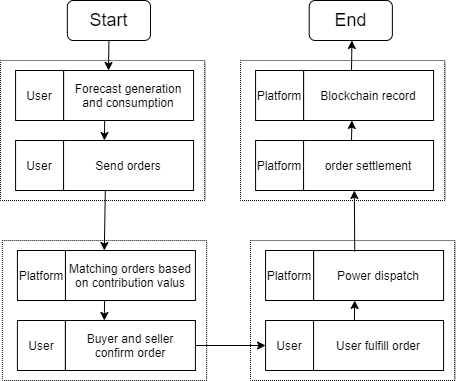}
    \caption{Blockchain-based P2P electricity trading process.}
    \label{fig:my_label}
\end{figure}

 The relationship between PoC consensus, blockchain and P2P energy trading is shown in Figure 3. We solve the problem of lack of trust between users in P2P energy transactions through blockchain technology. PoC consensus not only improves the efficiency of consensus and the fairness of node selection, but also encourages users to participate in energy transactions.

 \begin{figure}[h!]
    \centering
    \includegraphics[width=8.5cm,height=6cm]{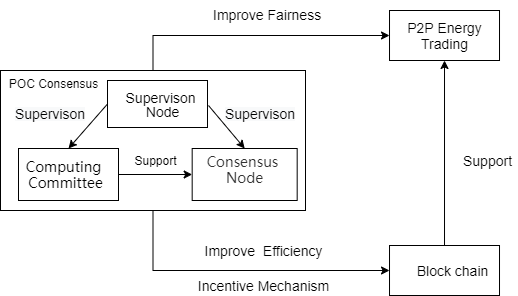}
    \caption{The interconnected relationship between PoC Consensus,blockchain,and P2P Energe Trading.}
    \label{fig:my_label}
\end{figure}

\section{ PoC consensus process}\label{}
\begin{figure}[h]
    \centering
    \includegraphics[width=8.5cm,height=5cm]{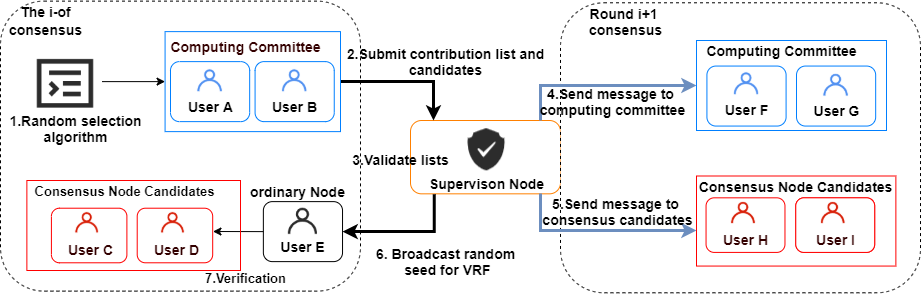}
    \caption{The Consensus Processes.}
    \label{fig:my_label}
\end{figure}

As the soul of the blockchain, the consensus algorithm ensures the consistency of transactions in the distributed system. A robust consensus algorithm can not only improve the efficiency of consensus, but also resist some malicious attacks. In this section, we designed a consensus algorithm suitable for microgrids, the specific design of the improved PoC consensus algorithm is presented in detail. Compared with the traditional consensus algorithm, our consensus algorithm not only need little computing resources, but also can attract users to participate in the blockchain through the contribution value as an incentive mechanism. The flow of PoC consensus algorithm is shown in Figure 4.

The VRF we use is an encryption scheme that maps the input to a verifiable pseudo-random output proposed by S. Micali et al.\cite{micali1999verifiable}. With the private key and the random seed as input, the pseudo-random number r and the corresponding proof $\pi$ are output, and other users can verify whether the pseudo-random number r is generated by the private key holder according to the random seed by using the public key corresponding to the private key, the proof $\pi$ and the random number seed, and the combination of blockchain and VRF can elect nodes randomly and securely.

\subsection{Random selection algorithm}\label{}
The number of  $N_{cs}$s and  $N_{cp}$s are adjusted by $N_{s}$ according to the total number of nodes in the blockchain. Setting more $N_{cs}$s and $N_{cp}$s will make the blockchain more secure, but it will reduce the efficiency of consensus. The $N_{cs}$s and $N_{cp}$s  for the first round of consensus are randomly selected by  $N_{s}$. Taking the i-th round of consensus as an example, each node in the computing committee ( $\Psi _{cp}^{i}$ ) of this round calculate the contribution values of all nodes during the previous consensus round to lists($L_{cb}$). Then each $N_{cp}$ in $\Psi _{cp}^{i}$ randomly selects a $N_{cp}$ and a $N_{cs}$ for the next round of consensus through Algorithm 1.

\begin{algorithm}
\SetKwInOut{Input}{Input}
\SetKwInOut{Output}{Output}
\Input{$L_{cb}$}
\Output{$N_{cp}$,$N_{cs}$}

$w_{sum} \leftarrow 0$\\
\For {each $weight_{x} \in L_{cb}$}{
    \If{$weight_{x}>0$}{
    $w_{sum}\leftarrow w_{sum}+weight_{x}$
    }
}
Random generate a weight $w_{r}(0<w_{r}<w_{sum})$\\
\For{each $weight_{x} $ in weights }{
    \If{$weight_{x}>w_{r}$}{
        $N_{cp} \leftarrow N_{x}$,Delete $w_{x}$ from weights\\
        break
    }
    }
Random generate a weight $w_{r}(0<w_{r}<w_{sum})$\\
\For{each $weight_{x} $ in weights }{
    \If{$weight_{x}>w_{r}$}{
          $N _{cs} \leftarrow N_{x}$, Delete $w_{x}$ from weights\\
        break
    }
    }
\caption{Random selection based on contribution value}
\end{algorithm}

\subsection{Submit contribution list}\label{}
The $L_{cb}$ will be obtained in descending order, and  the $L_{cb}$ format is expressed as follows:
\begin{equation}
<node_{id},pk,value,history,weight>
\end{equation}
where $node_{id}$ represents the node id, pk represents the public key of the node, value represents the accumulated contribution value of the node, history represents the historical contribution value record of the node, and weight represents the weight of the node calculated based on the contribution value.

$N_{cp}s$ submit  $L_{cb}s$ to $N_{s}$ for verification. Take computing node A as an example, the message format sent to the supervising node is expressed as follows:
\begin{equation}
<node_{A},L_{A},\pi_{A},R_{A},N_{cp}^{A}, N_{cs}^{A} >
\end{equation}
where $node_{A}$ represents the ID of A, $L_{A}$represents the list of contribution values submitted by A, $\pi_{A}$ represents the proof $\pi$ generated by random seed($R_{i-1}^{cp}$) through VRF, $R_{A}$ represents the pseudo-random number, $N_{cp}^{A}$ represents the next round of computation node randomly selected  by A, $N_{cs}^{A} $ represents the next round of consensus node. All information is signed by the sender and encrypted with the public key of the recipient  for security purposes.

\subsection{Validate lists}\label{}
\begin{algorithm}
\SetKwInOut{Input}{Input}
\SetKwInOut{Output}{Output}
\Input{$M_{x}$,n}
\Output{$L,\Psi_{cs}^{i+1},\Psi_{cp}^{i+1}$}
\BlankLine
Initialize $\Psi=\phi, \Psi_{cs}^{i+1}=\phi,\Psi_{cp}^{i+1}=\phi$
\BlankLine
\While{Receive message from $N_cp$}{
\If{$VRF_{ver}(\pi_{Node_{x}},R_{i-1}^{cp},PK_{x}$)=Accept}{
        $\Psi  = \{M_{x}\}\cup \Psi$
    }
\If{$\Psi>=n$}{
counts $\gets$ empty dictionary\\
  \For{$item$ in $ \Psi$}{
    \If{$item.L$ in $counts$}{$counts[item] \gets counts[item] + 1$}
    \Else{ $counts[item] \gets 1$}
  }
  \For{$item, count$ in $counts$}{
  \If{$count > n$}{
  \If{$ Node_{cs}^{item.weight} >0 $ and $ Node_{cp}^{item.weight}>0$}{  
  $\Psi_{cs}^{i+1}= \Psi_{cs}^{i+1}\cup Node_{cs}^{item.id} $,
  $\Psi_{cp}^{i+1}= \Psi_{cp}^{i+1}\cup Node_{cp}^{item.id} $}
  $L = item.L $
  }
  }
    }
}

\caption{Verified contribution list }
\end{algorithm}

$N_{s}$ receives message and verifies the list by Algorithm 2. In the algorithm, $M_{x}$ is the message sent by the $N_{cp}$, n is a hyperparameter that represents the threshold value, and the value of n is set according to the number of computing committee nodes in the network. We recommend that n be set to one half of the number of computing committee. When the number of correct is grater than  n, we consider that there is agreement on the results of the list. Meanwhile the $N_{s}$ also determines whether the candidate node submitted by the $N_{cp}$ is reasonable. With Algorithm 2, we get the list of contribution values $L$, the consensus candidate nodes $\Psi_{cs}^{i+1}$ and computing committee $\Psi_{cp}^{i+1}$ for the next round. If the number of $\Psi_{cs}^{i+1}$  and $\Psi_{cp}^{i+1}$ does not reach the required number, $N_{s}$ executes Algorithm 1 for replenishment. Malicious nodes do not have random number seeds and cannot pass validation even if the list is submitted. If the list submitted by $N_{cp}$ is not verified by $N_{s}$, then this node will be considered a malicious node and will not be selected as a member of the consensus committee.

\subsection{Send message to computing committee}\label{}
$N_{s}$ send messages to the corresponding nodes according to the  $\Psi_{cp}^{i+1}$. The messages format  sent to the  $\Psi _{cp}^{i+1}$ are expressed as follows:
\begin{equation}
<R_{i}^{cp},L >
\end{equation}
where $R_{i}^{cp}$ represents the seed of the random number generated in this round, $L$ represents the list of contribution values.

 If L and $R_{i}$ are received from the supervising node, that node is a $N_{cp}$ for the next round. The node that receives the message will generate the corresponding proof $\pi$ and pseudo-random number based on the random seed $R_{i}^{cp}$.

\subsection{ Send message to consensus candidates}\label{}
$N_{s}$ send messages to the corresponding nodes according to the  $\Psi_{cs}^{i+1}$. The messages format sent to the  $\Psi _{cs}^{i+1}$ is expressed as follows:
\begin{equation}
<R_{i}^{cs} >
\end{equation}
where $R_{i}^{cs}$ represents the seed of the random number generated in this round.
If a random seed is received from a supervisory node, that node  is a $N_{cs}$. The node that receives the message will generates the corresponding proof $\pi$ and pseudo-random number based on the random seed $R_{i}^{cs}$.

\subsection{ Broadcast random seed}\label{}
We set a trading time for each round, and when the time limit is reached or the number of trades exceeds the threshold. $N_{s}$ will broadcast a message. The message format broadcast by the supervising node is as follow:
\begin{equation}
<R_{i-1}^{cs},\Psi _{cs}^{i},\Psi _{cp}^{i}>
\end{equation}
where $R_{i-1}^{cs}$ is the random number seed generated in the previous round, $\Psi _{cs}^{i}$ are the consensus candidates node for this round, $\Psi _{cp}^{i}$ are the computing committee for this round.

\subsection{ Verification blocks}\label{}
We set a window period T to ensure enough time to complete consensus. $N_{cs}$s stop packing and broadcast their own block. $N_{o}$s verify the blocks based on $R_{i-1}^{cs}$, and the block with the largest random number among $N_{cs}$s is used as the block on the chain. If the packaged transaction in this block does not get the consent of half of the nodes, then the new block is complemented in the order of random number size.








\section{Contribution value related design}\label{}
 The rationality of contribution value design determines whether users can be motivated. We quantifies the user's contribution to the blockchain and gives the corresponding specific contribution value based on the user's transaction behavior and participation in consensus tasks in the network. We divide the contribution value into two components: the contribution of energy trading(CE), which is visible to all users, and the consensus task contribution value (CC), which is known only to the nodes participating in the consensus task. In the figure below, we show the assessment index of contribution value. Through the contribution value mechanism, high-quality users are selected to allow them to undertake more tasks in the blockchain, which can not only ensure the safe operation of the blockchain network, but also motivate users and attract users to join the microgrid blockchain.
\begin{figure}[h]
    \centering
    \includegraphics[width=8.5cm,height=6cm]{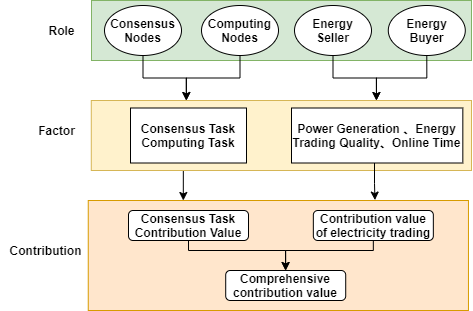}
    \caption{Schematic Diagram of Contribution Value Classification. }
    \label{fig:my_label}
\end{figure}
\subsection*{5.1 The contribution of energy trading }
The assessment indexes selected for CE are power generation, energy trading quality and nodes stable online time.

\subsubsection*{5.1.1 Power generation contribution }\label{}
The power generation of the equipment is the first step in the energy trading of the microgrid. Users can participate in the transaction only if they have excess energy, so we hope to encourage users to actively participate in power generation under conditions. The contribution of power generation is calculated as follow:
\begin{equation}
PGC= {\textstyle \sum_{1}^{n}} (\alpha _{1}*P_{n}) 
\end{equation}
where n indicates the number of distributed generation units owned by the customer, $\alpha_{1}$ indicates the contribution value that can be obtained per kW of electricity generated by the device, $P_{n}$ indicates the actual amount of power emitted by the device.

\subsubsection*{5.1.2 Energy trading quality }
We introduce transaction quality $TQ$ to quantify the transaction quality of an electricity transaction.The formula for the transaction quality (TQ) is shown as follow:
\begin{equation}
TQ=1-(\frac{P_{order} -P_{real}}{P_{order}})
\end{equation}
where $p_{real}$ denotes the actual electricity delivered by the generation facility, $p_{order}$ denotes the electricity agreed upon by the buyer and seller in the order.

Meanwhile, in order to prevent users from maliciously swiping contribution value by intentionally making small power transactions to increase the number of transactions. We introduce the mechanism of diminishing returns. The user 's multiple transactions in a round of consensus process will reduce the contribution of subsequent transaction acquisition. But in the actual situation, the user needs to conduct multiple power transactions. We consider that this part of the contribution value has been considered in the power generation contribution value stage. The mechanism of diminishing returns will not affect user participation. This stage is mainly to encourage users to conduct normal power transactions. The user's electricity transaction contribution value (ETC) calculation formula is defined as follow:
\begin{equation}
ETC= {\textstyle \sum_{1}^{n}}\frac{\alpha_{2}*TQ_{n}}{n^2} 
\end{equation}
where n represents the number of transactions initiated by the user during the previous round of consensus. $TQ_{n}$ represents the transaction quality of the nth transaction, and $\alpha_{2}$ represents the contribution value that can be obtained by completing an electricity transaction.

\subsubsection*{5.1.3 Stable online time }\label{}
 If the number of nodes in a blockchain network is unstable, the network is unsafe and untrusted. Only the sufficient number of nodes can quickly verify and forward information and resist the attack of malicious nodes.The formula for the node stable online time contribution (SOT) is shown as follow:
\begin{equation}
SOT = \alpha_{3}*(T_{off}-T_{on})
\end{equation}
where $\alpha_{3}$ denotes the weight of online time in the contribution value of power transactions, $T_{on}$ denotes the last time on-line, and $T_{off}$ denotes the latest time to go offline.

The contribution value of energy trading is calculated as follow:
\begin{equation}
CE = PGC+ETC+SOT
\end{equation}

\subsection*{5.2 The contribution of consensus  }
Similarly, in order to encourage nodes to undertake the consensus task of the blockchain, we will reward nodes that correctly complete the task. We screen out high-quality nodes through contribution values to improve the efficiency of blockchain consensus. The Contribution of consensus is calculated as follow:
\begin{equation}
CC = \theta_{1} {\textstyle \sum_{1}^{n}\alpha _{4}}+\theta_{2} {\textstyle \sum_{1}^{m}\alpha _{5}}
\end{equation}
where $\alpha_{4}$ represents the contribution value that can be obtained as a computing node, n represents the number of nodes in the blockchain network, $\alpha_{5}$ represents the contribution value obtained as a consensus node, and m represents the number of packaged transactions as a consensus node.If this node acts as a computing node, $\theta_{1}$ takes 1 and $\theta_{2}$ takes 0. If the node is a consensus node, $\theta_{1}$ takes 0 and $\theta_{2}$ takes 1.

The total contribution of a node is calculated as follow:
\begin{equation}
N_{C} = CE+CC
\end{equation}

\subsection*{5.3 The weight based on contribution value}
Each time a node participates in a round of consensus, it will get a contribution value. We calculate the weight of node based on these contribution values. The weight based on contribution value of a node is calculated as follow:
\begin{equation}
W = \left\{\begin{matrix} \frac{1}{\sqrt{ {\textstyle \sum_{i=1}^{n}} (N_{c}^{i}-\mu )^{2}/n } }  ,n>1
 \\\frac{1}{N_{C}^{i} },n=1
\end{matrix}\right.
\end{equation}
where $N_{c}^{i}$ denotes the contribution value of the i th round of the node, n represents the number of nodes participating in the consensus, $\mu$ represents the average contribution value of the node.

\section{Security Analysis}
The proposed PoC consensus microgrid blockchain platform is a public blockchain system without CA endorsement. Because there is no access mechanism, anyone can join the platform, so there may be malicious users to launch various attacks on blockchain network, which will lead to security risks to the system. Now we will analyze the security of common attacks and problems of public blockchain.

\subsection*{6.1 Multi-node collusive attack}
In the PoC consensus, if the nodes in the $\Psi_{cs}$ collude, the $L$ will be manipulated. We only randomize the nodes within the set, so that the random meaning is lost, and the blockchain also loses fairness. However, in our scheme, the nodes in $\Psi_{cs}$  only know their identity and do not know who the other $N_{cp}$s are. The L is carried out through the $N_{s}$, and there is no possibility of collusion attack.
\subsection*{6.2 Node monopoly problem}
Monopoly has always been a problem that cannot be ignored in the PoC consensus mechanism. Personal distributed power generation equipment is impossible to compare with power plants in terms of power generation and power transactions. In this paper, we first reset the contribution value of the consensus node.Secondly, the mechanism of diminishing contribution income is added to limit the income of power plants. Finally, the contribution value is further divided into two contribution values : CE and CC. When calculating the node weight, the two contribution values are calculated separately. The high contribution value of CE does not mean that the CC is high. If the malicious node attacks the node with the highest CE , the attacked node is not necessarily the node participating in the consensus task, which improves the stability of the blockchain to a certain extent.

\section{Experiment}
 In this section, we evaluate our proposed consensus algorithms through a series of simulation experiments. Through simulation, we can evaluate the feasibility and security of our scheme and the efficiency of consensus. And through simulation experiments, we verify whether the contribution value algorithm we designed is reasonable.
 
 We implement a complete consensus process and contribution value calculation process through GoLang. The experiment is carried out in the simulation environment, using random numbers to simulate power generation. All tests below are done on windows 11, and the hardware we used is a mainframe with a 3.2 GHz Intel Core i7 processor and 16GB RAM.

\subsection{Evaluation of PoC and contribution value }
 One of the characteristics of blockchain is deneutralization. If a node continuously acts as an accounting node, then the blockchain system will not be trusted. In our proposed PoC consensus algorithm, we use VRF for random selection. After the node becomes a consensus node, the contribution value of the node will be emptied to prevent the monopoly of the node. In theory, the rights of accounting should be evenly distributed between nodes. Figure 6 shows the distribution of accounting nodes during the 100 rounds of consensus. In the 100 rounds of consensus, all nodes have the opportunity to become accounting nodes, and there has been no monopoly of nodes. 
\begin{figure}[h]
    \centering
    \includegraphics[width=8cm,height=6cm]{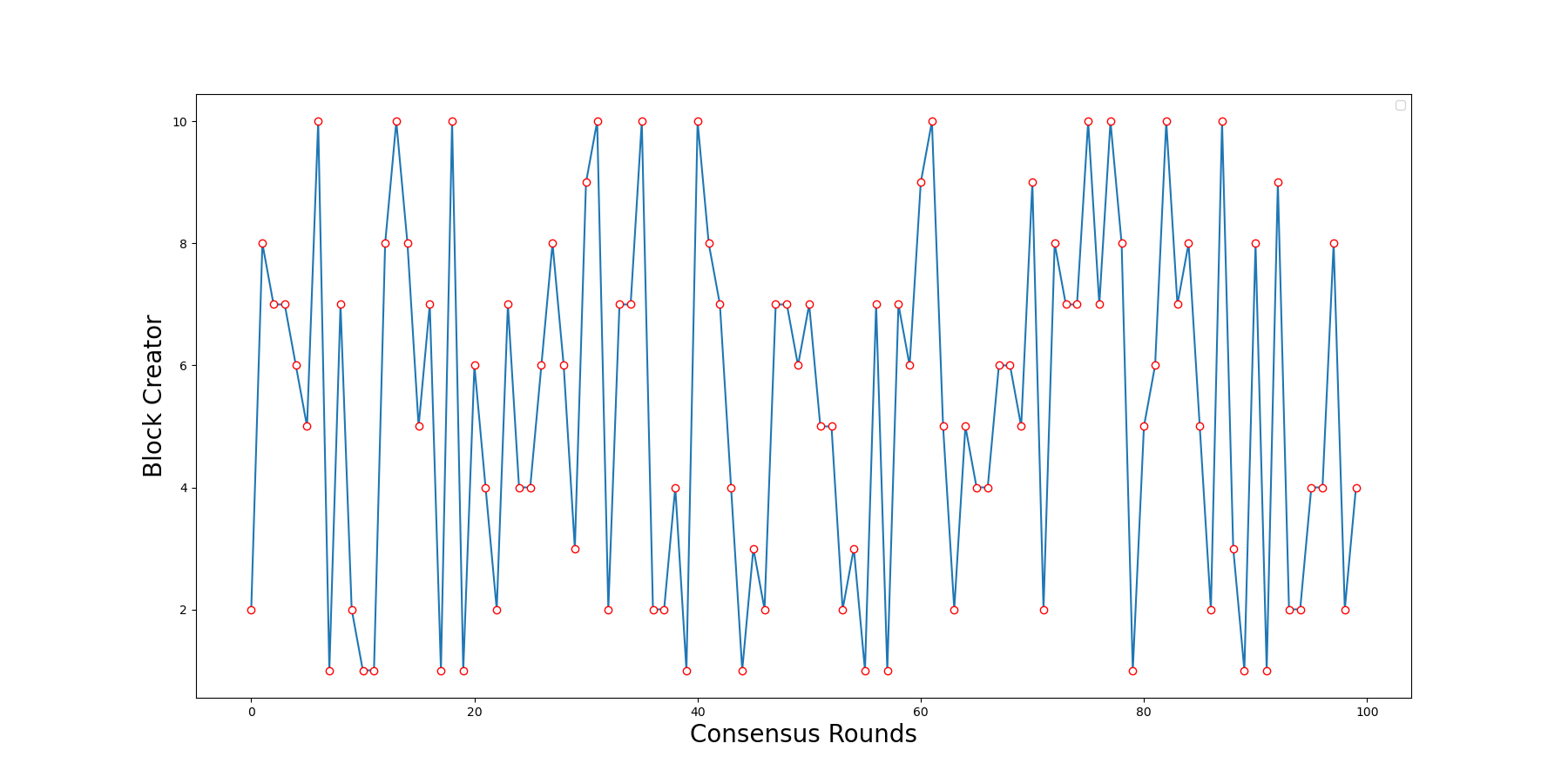}
    \caption{Distribution of Block Generating Nodes. }
    \label{fig:my_label}
\end{figure}

Figure 7 shows the selection of nodes in the 100-round consensus. We assume that each node is honest and set 10 nodes, where the number of members of the calculation committee is set to 4 in each round. Ideally, in 100 rounds of consensus, each node should be elected 10 consensus nodes and 40 computing nodes on average. In the experiment, the distribution of nodes indeed satisfies the distribution.
\begin{figure}[h]
    \centering
    \includegraphics[width=8cm,height=6cm]{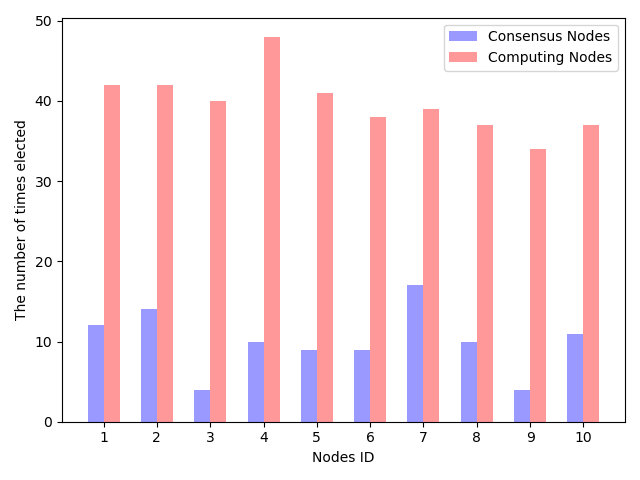}
    \caption{Distribution of Node Elections. }
    \label{fig:my_label}
\end{figure}

As shown in Figure 8, we verify the effect of diminishing returns on the contribution value of electricity trading designed by Formula (8). It can be found that when the user completes the power transaction multiple times in a round of consensus, its revenue will decrease exponentially.

\begin{figure}[h]
    \centering
    \includegraphics[width=8cm,height=6cm]{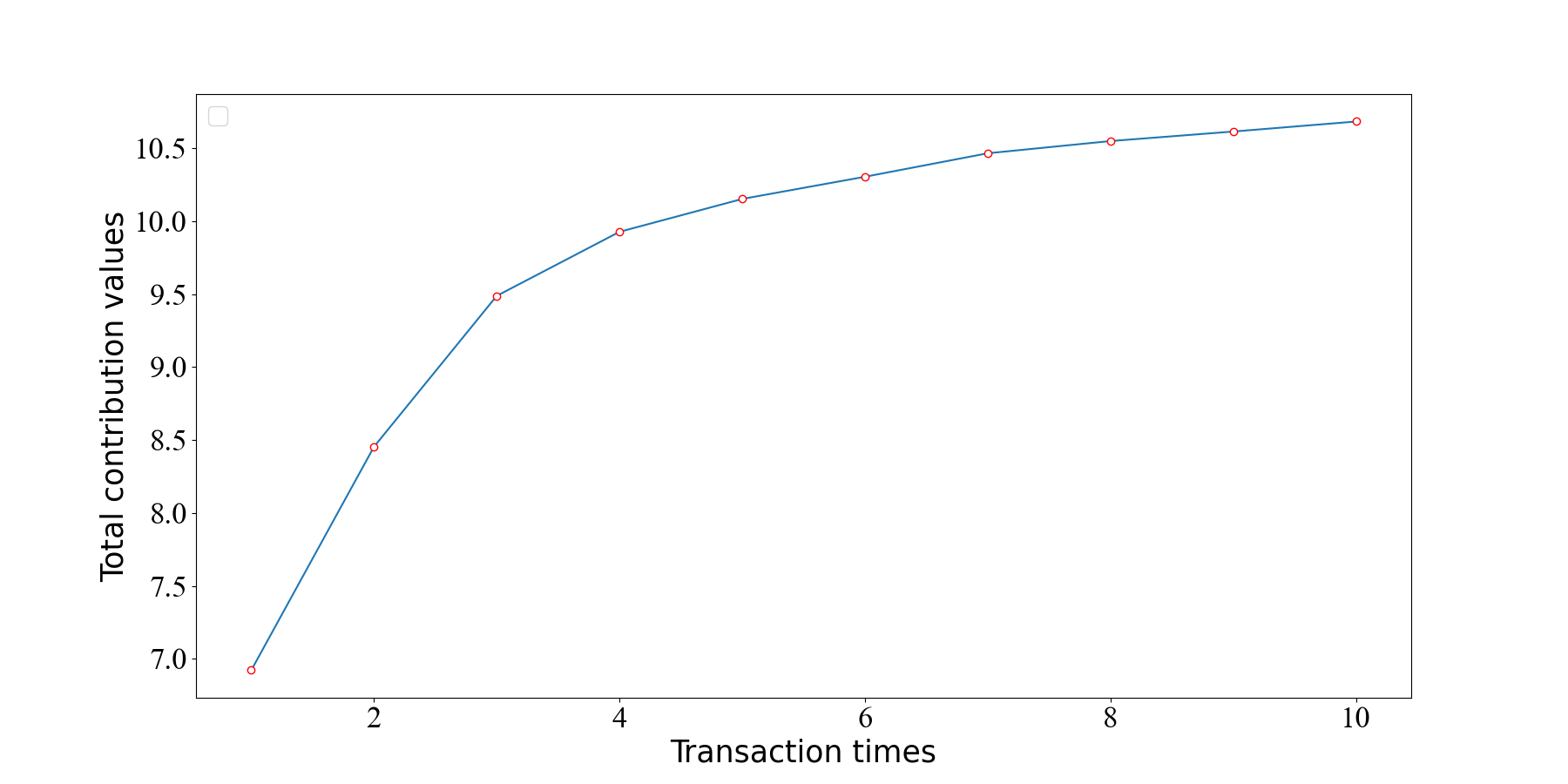}
    \caption{Contribution Value Income Diminishing Effect. }
    \label{fig:my_label}
\end{figure}
As shown in Figure 9, we track the change of contribution value of an honest node. It can be found that if a node is honest, its weight can be steadily improved, but it does not mean that the node with high contribution value has high weight of consensus node.Because the weight of the node is not only related to the contribution value of the node, but also related to the historical situation of all nodes in the blockchain network.

\begin{figure}[h]
    \centering
    \includegraphics[width=8cm,height=6cm]{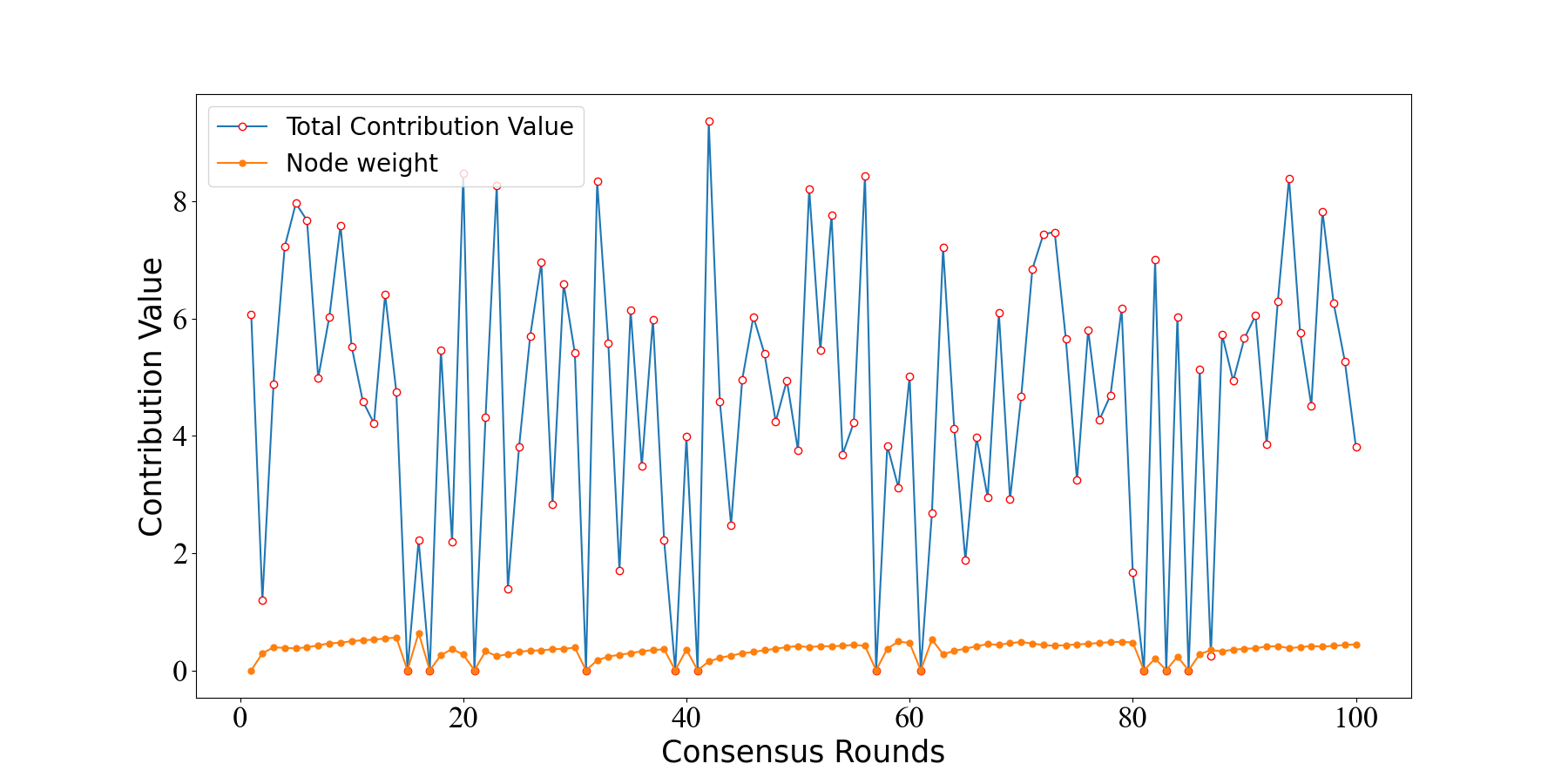}
    \caption{Distribution of Node Contribution Value And Weight. }
    \label{fig:my_label}
\end{figure}

We simulate the presence of malicious nodes in the node. In the blockchain system composed of 10 nodes, we simulate the presence of one and two malicious nodes respectively. Once the node has a submitted error list or the block verification does not pass, it will be identified as a malicious node. In our design, the contribution value of the malicious node will not be calculated. In theory, they will not be selected as consensus nodes or computing nodes in subsequent consensus rounds. As shown in Figure 10, the election of consensus nodes in 100 rounds of consensus, malicious nodes have not become consensus nodes after the first round of evil.

\begin{figure}
        \center
        \scriptsize
        \begin{tabular}{cc}
                \includegraphics[width=8cm,height=6cm]{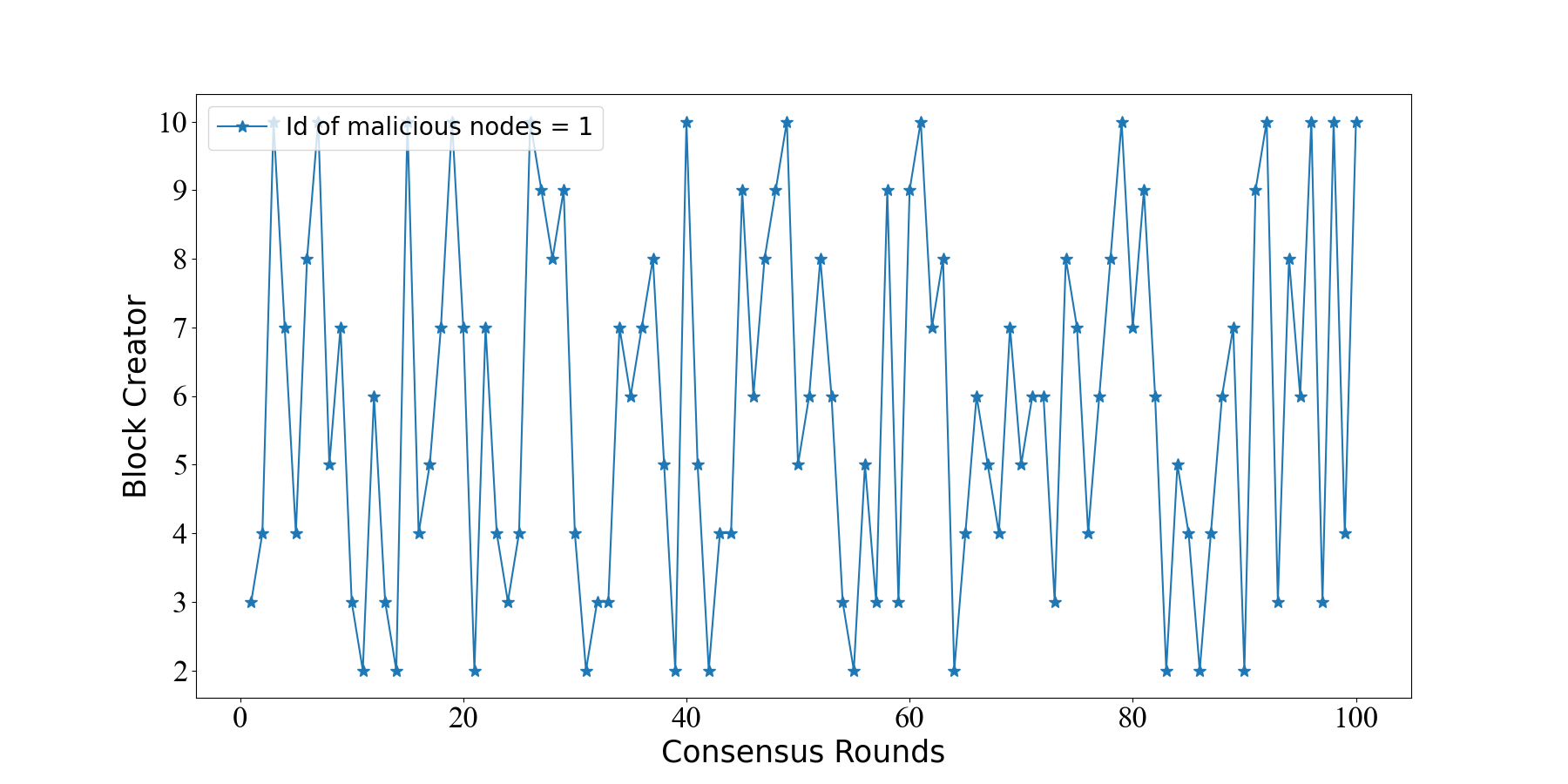} &    
                \includegraphics[width=8cm,height=6cm]{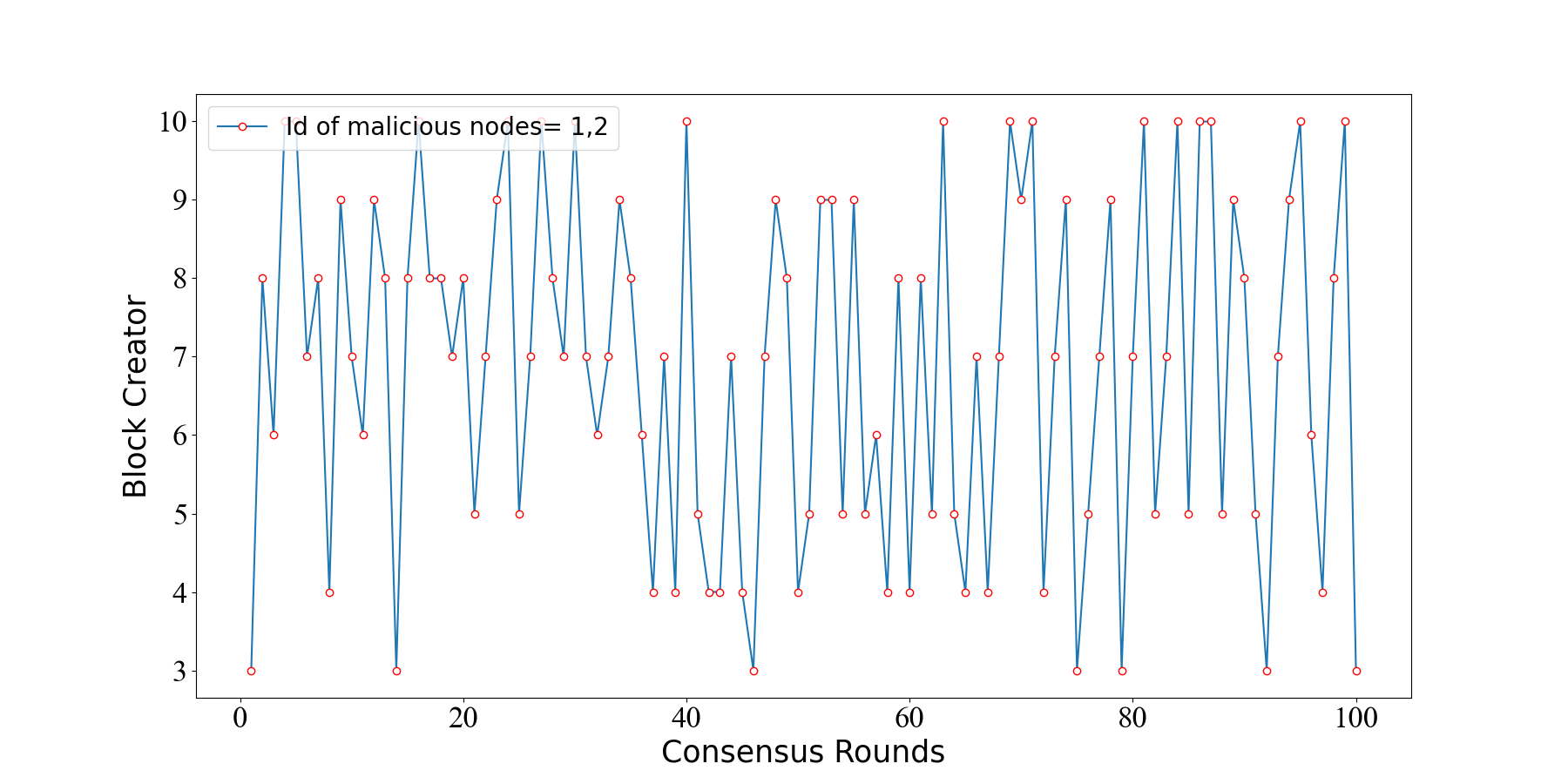}        \\
                (a) The number of malious node is 1. & (b) The number of malious nodes is 2. \\
        \end{tabular}
        \caption{Distribution of Block Generating Nodes}
        \label{figure}
        \vspace{-0.5em}
\end{figure}

\subsection{Comparison of PoC and PoW efficiencies }
Figure 11 shows the block generating time of PoC and PoW. PoW is the most classic consensus algorithm in the blockchain. Affected by machine computing power, the time to find a random number that meets the difficulty requirements in each round is uncertain. Therefore, the time to generate a block in each round of PoW is unstable. In our scheme, computing nodes in the consensus committee calculate the contribution value during the previous round of consensus, and randomly selects the next round of consensus committee based on the weight of the contribution value. Compared to calculating meaningless puzzles, we need very little time to calculate the value of the contribution.

\begin{figure}[h]
    \centering
    \includegraphics[width=8cm,height=6cm]{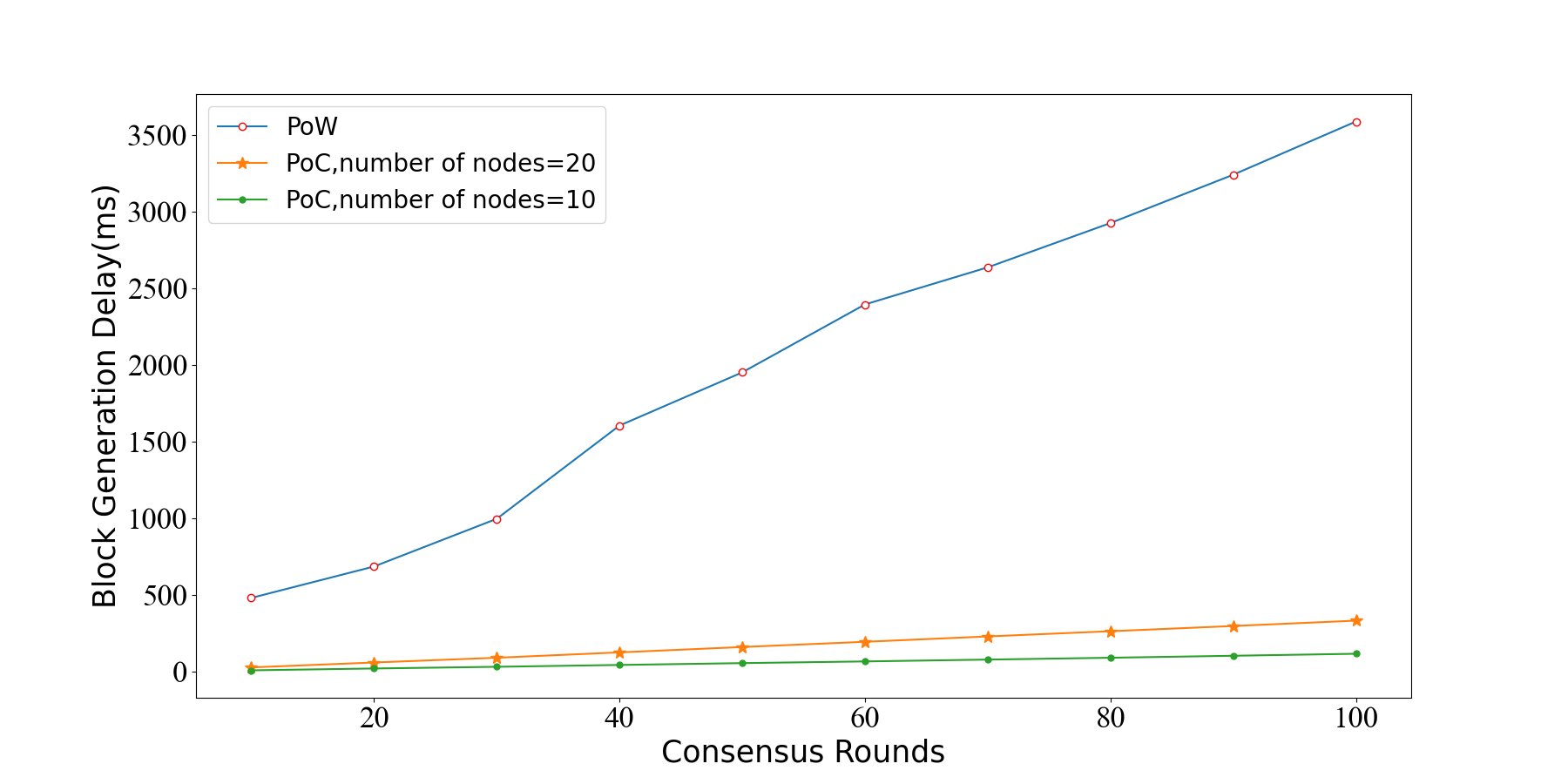}
    \caption{Comparison of Efficiency on New Block Generation }
    \label{fig:my_label}
\end{figure}

\section{Conclusion}
In this study, we propose a microgrid trading framework based on PoC consensus. Our blockchain energy trading framework does not rely on cryptocurrencies and we design a contribution value mechanism suitable for microgrids. The contribution value is classified into two parts : power transaction and blockchain consensus, which not only ensures the fairness of power transaction, but also improves the efficiency of blockchain consensus. Ordinary nodes can only view the contribution value of electricity transactions, and only the selected nodes can know the complete list of contribution values. Our consensus algorithm will randomly select nodes based on the weight of the contribution value to form a consensus committee. The results of experimental simulation show that the efficiency of our consensus algorithm is significantly improved compared with the traditional consensus method. The contribution value of the node does not appear monopoly phenomenon, each node has the opportunity to become a consensus node. The existence of supervisory node makes malicious nodes not participate in the consensus task, which ensures the stability and correctness of the blockchain network.

In the future work, we will study alternative methods of supervision node to achieve a more de-neutralization scheme. We will constantly optimize the calculation of contribution value and expand the benefits of users with high contribution value, such as giving users with high contribution value more benefits when pricing electricity.
 
\section*{Acknowledgements}
The research was supported in part by  the Guangxi Science and Technology Major Project (No. AA22068070), the National Natural Science Foundation of China (Nos. 62166004,U21A20474), the Key Lab of Education Blockchain and Intelligent Technology, the Center for Applied Mathematics of Guangxi, the Guangxi "Bagui Scholar" Teams for Innovation and Research Project, the Guangxi Talent Highland Project of Big Data Intelligence and Application, the Guangxi Collaborative Center of Multisource Information Integration and Intelligent Processing.
\bibliographystyle{cas-model2-names}

\bibliography{cas-refs}

\bio{}
\endbio

\endbio

\end{document}